\documentclass[letterpaper, 10 pt, conference]{ieeeconf}
\IEEEoverridecommandlockouts
\usepackage{cite, amsmath, amssymb, amsfonts, algorithmic, graphicx, textcomp, xcolor, epsfig, amsfonts, bm, amstext, booktabs, verbatim, color, makecell, array}

\newcommand{\fullname}{Multi-View Attention Learning}
\newcommand{\shortname}{MuVAL}

\def\BibTeX{{\rm B\kern-.05em{\sc i\kern-.025em b}\kern-.08em
    T\kern-.1667em\lower.7ex\hbox{E}\kern-.125emX}}

\title{\LARGE \bf 
Multi-View Attention Learning for Residual Disease Prediction of Ovarian Cancer
}

\author{
Xiangneng Gao$^{1}$, Shulan Ruan$^{2}$, Jun Shi$^{2}$, Guoqing Hu$^{3}$, and Wei Wei$^{3}$
\thanks{$^{1}$Xiangneng Gao is with the Department of Computer Science and Engineering, 
        Southern University of Science and Technology, Shenzhen, 518055, China~(\tt\small xiangnenggao@gmail.com)}%
\thanks{$^{2}$Shulan Ruan is the corresponding author.
        Shulan Ruan and Jun Shi are with the School of Computer Science and Technology, 
        University of Science and Technology of China, Hefei, 230026, China~(\tt\small slruan@mail.ustc.edu.cn, shijun18@mail.ustc.edu.cn)}%
\thanks{$^{3}$Guoqing Hu and Wei Wei are with the Department of Radiology, 
        the First Affiliated Hospital of USTC, Hefei, 230001, China~(\tt\small huguoqing9792@163.com, weiweill@126.com)}%
}

\begin{document}

\maketitle
\thispagestyle{empty}
\pagestyle{empty}

\begin{abstract}
In the treatment of ovarian cancer, precise residual disease prediction is significant for clinical and surgical decision-making.
However, traditional methods are either invasive~(e.g., laparoscopy) or time-consuming~(e.g., manual analysis).
Recently, deep learning methods make many efforts in automatic analysis of medical images.
Despite the remarkable progress, most of them underestimated the importance of 3D image information of disease, which might brings a limited performance for residual disease prediction, especially in small-scale datasets.
To this end, in this paper,
we propose a novel Multi-View Attention Learning~(MuVAL) method for residual disease prediction,
which focuses on the comprehensive learning of 3D Computed Tomography (CT) images in a multi-view manner.
Specifically, we first obtain multi-view of 3D CT images from transverse, coronal and sagittal views.
To better represent the image features in a multi-view manner, we further leverage attention mechanism to help find the more relevant slices in each view. 
Extensive experiments on a dataset of 111 patients show that our method outperforms existing deep-learning methods. 

\end{abstract}
\section{Introduction}
\label{sec:introduction}

Ovarian cancer is one of the cancers with the highest incidence and mortality rates among women worldwide~\cite{sung2021global}. 
The general treatment is cytoreductive surgery; however, residual disease from an incompletely successful surgery may cause extensive metastasis, increasing the risk of postoperative recurrence and mortality~\cite{chen2021ct}.
Therefore, it is crucial to accurately predict the residual disease after cytoreductive surgery to make clinical surgical decisions.
In this paper, we refer to the objective as \emph{residual disease prediction of ovarian cancer.}

As far as we are concerned, large efforts have been made by researchers from various fields, such as medical and computer science, to accurately predict the residual disease of ovarian cancer.
In earlier stages, laparoscopy, a kind of invasive examination, was applied to provide a visual and systematic view of the whole abdomen \cite{testa2012ultrasound, van2019laparoscopy}.
Later, Li et al. \cite{li2021noninvasive} employed feature engineering and statistical method to analyze clinical and medical imaging data obtained by noninvasive examination.
With the successful accomplishment of deep learning, significant progress has been made in medical image analysis.
Wang et al. \cite{wang2019deep}, proposed to leverage a CNN-based framework to extract deep image features for recurrence prediction of ovarian cancer.
\begin{figure}
    \centering
    \includegraphics[height=45mm, width=65mm]{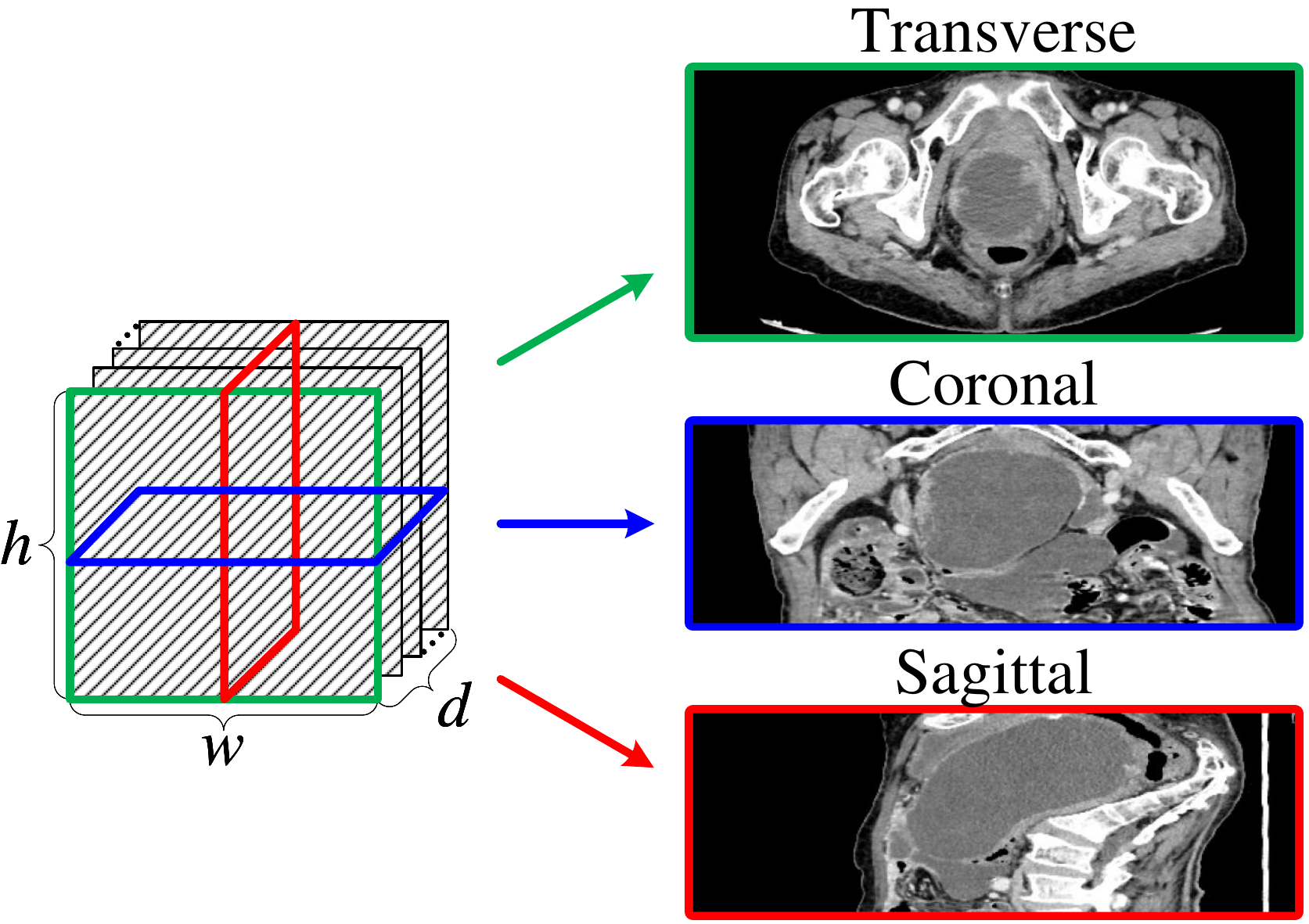}
    \caption{Multi-view of a 3D CT image.}
    \label{f:MuVAL_sample}
\end{figure}

However, traditional invasive methods sometimes naturally carry an additional risk of metastasis.
Furthermore, feature engineering and statistical methods often require manual selection and design of features by knowledgeable medical professionals, which is both time-consuming and labor-intensive.
In addition, despite the remarkable progress of deep-learning technology in medical image analysis,
most of the current deep-learning methods underestimate the importance of 3D image information contained in Computed Tomography (CT) or other medical images, which may bring a limited performance for residual disease prediction.

As a matter of fact, a 3D CT image of a patient's entire disease region is composed of a stack of $d$ 2D CT slices, each with a resolution of ($h, w$). 
According to the size of the disease region, $d$ is usually between 10 and 100.
Traditionally, the 2D CT slices used in the previous works only included the transverse view (in green), as shown in Fig~\ref{f:MuVAL_sample}.
However, multi-view features facilitate a more comprehensive and accurate observation of 3D images~\cite{wildeman20092d, wu2020deep}.
Thus, by fully utilizing multi-view features such as the coronal view (in blue) and the sagittal view (in red), more specific information about the disease region can be obtained. 
Although the value of multi-view features has been proved, there are still many challenges in better utilizing multi-view features in 3D CT images. 
The difficulty of image representation is compounded by factors such as the shape, size, texture, invasion, and metastasis of the disease, all of which can affect the outcome of surgery. Additionally, the informativeness of 2D CT slices varies, with some slices reflecting the residual disease situation better than others. 
Further, differences in cancer subtypes and surgical teams introduce differences in surgical procedures, and specific surgical teams and cancer subtypes will limit the scale of available data.

To this end, in this paper, we propose a novel Multi-View Attention Learning~(MuVAL), an end-to-end deep-learning method for residual disease prediction of ovarian cancer. 
Specifically, we first obtain transverse, coronal, and sagittal views of 3D CT images to create multi-view images.
To better represent the image features, we leverage Squeeze-and-Excitation~(SE) \cite{he2016deep} module to help find the more relevant slices in each view.  
In the backbone, we apply Med3d \cite{chen2019med3d}, a pre-trained ResNet \cite{he2016deep} for 3D medical image, to deal with the small-scale dataset.
Extensive experiments and multiple metrics demonstrate the superiority of our proposed method.
As an emphasis, the main contributions of our work can be concluded as follow: 
\begin{itemize}
    \item We observe the great potential of utilizing multi-view of 3D CT images from transverse, coronal, and sagittal views, and propose to leverage the multi-view features to enhance the learning for residual disease prediction of ovarian cancer.
    \item We propose a novel Multi-View Attention Learning (MuVAL) method for residual disease prediction of ovarian cancer, which focuses on the comprehensive learning of 3D CT images in a multi-view manner.
    \item Extensive experiments and multiple metrics demonstrate the superiority and rationality of our proposed method compared with the baseline methods.
\end{itemize}
\section{Related Work}
\label{s:related_work}

Researchers have put a lot of effort into predicting residual disease in ovarian cancer\cite{chi2006optimal, ferrandina2009role, shim2015nomogram}, including manually analyzing or designing models based on data generated from various preoperative examinations.
In the early stages, doctors systematically observe the whole abdomen through laparoscopy to predict the residual disease\cite{testa2012ultrasound, van2019laparoscopy}. However, Chi et al.\cite{chi2004ten} have highlighted the potential risks of invasive examination and associated complications.


In recent years, a growing number of researchers have shifted their focus to non-invasive methods for predicting residual disease in ovarian cancer. Many studies have employed multivariate analysis using preoperative clinical data, including physical, hematological, genetic, and other patient indicators. 
For instance, Laios et al.\cite{laios2020predicting, laios2020247} used age, Body Mass Index (BMI), surgery timing, and other variables to develop a k-Nearest Neighbor (k-NN) model for prediction. Meanwhile, Berchuck et al.\cite{berchuck2004prediction} examined the association of residual disease with genes and screened 120 genes to construct a predictive model.
Another approach proposed by Shah et al.\cite{shah2018combining} is the use of serum CA-125 levels and serum microRNA levels as predictive markers through stepwise logistic regression analysis. 
Moreover, in recent years, preoperative medical image analysis with deep learning methods~\cite{shi2021darnet} has become an increasingly influential method for predicting residual disease. Medical image data, such as CT and Magnetic Resonance Imaging (MRI), are visual and informative.
For instance, Barber et al.\cite{barber2021natural} utilized preoperative CT reports to develop a Natural Language Processing (NLP) model.
Li et al.\cite{li2021noninvasive} employed Radiomics\cite{gillies2016radiomics} features extracted from MRI with clinical data. 
Last but not least, Peritoneal Carcinomatosis Index (PCI), an imaging indicator based on disease size and location within the peritoneal cavity, has been widely applied to predicting residual disease \cite{llueca2018prediction, engbersen2019mri, jonsdottir2021peritoneal, li2022diffusion}.


Although no studies have been conducted specifically to predict residual disease in ovarian cancer using deep-learning methods, such methods have been applied to related tasks, such as diagnosis\cite{saida2022diagnosing}, histotype classification\cite{farahani2022deep}, and recurrence prediction\cite{wang2019deep}. 
Notably, these studies have demonstrated that deep-learning methods and medical image analysis can match or even outperform experienced doctors, highlighting their potential for clinical use.


\section{Problem Statement and Method}
\label{s:method}

\subsection{Problem Statement}

In this section, we formulate the task as a supervised binary classification problem.
Given a 3D CT image $\bm{I}\in\mathbb{R}^{d \times h \times w}$ of a patient,
our goal is to learn a classifier $\xi$ which can precisely predict the residual disease $y$,
with $y = \xi(\bm{I})$.
No residual disease is known as R0 and vice versa as non-R0 in medical terms.
In this paper, we denote $y=1$ for R0 and $y=0$ for non-R0.

\subsection{Overall Architecture}
\begin{figure*}
    \centering
    \includegraphics[width=\textwidth]{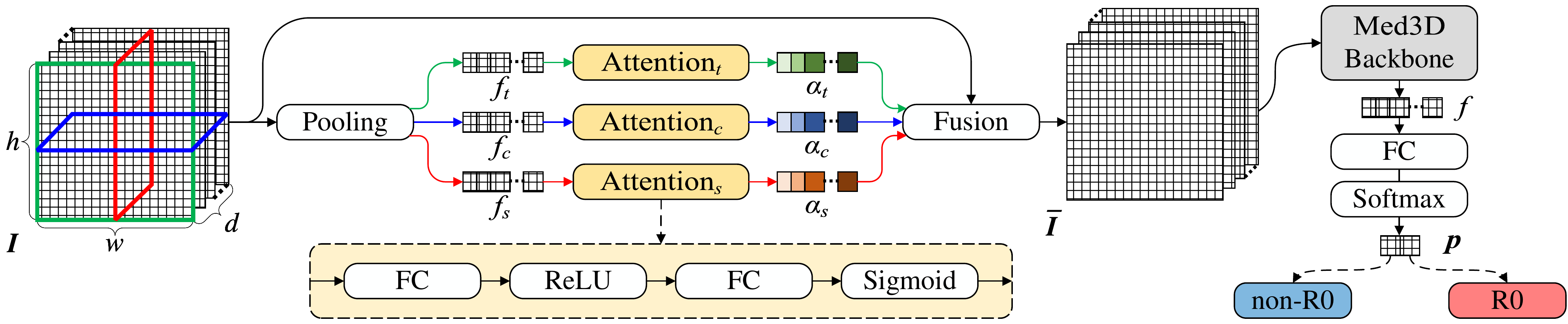}
    \caption{The architecture of MuVAL.}
     \label{f:framework}
\end{figure*}

The overall architecture of our proposed method is shown in Fig.~\ref{f:framework}.
Firstly, we use pooling to create multi-view slice features from the various dimensions of 3D CT images.
Then, we develop a~\fullname~(\shortname) to learn more valuable image features under different views with an attention mechanism.
Finally, we integrate the learned multi-view features and employ Med3D to help the final label prediction.
In the following part, we will introduce~\shortname~in detail.

\subsection{Multi-View Attention Learning}
As shown in Fig.~\ref{f:MuVAL_sample}, 
the raw CT image has the natural characteristics of a 3D structure.
Observing from different views can obtain different features and thus enhance the representation of 3D CT images.
For a 3D CT image $\bm{I}\in\mathbb{R}^{d\times h \times w}$,
we consider the transverse view as the default view and augment it with the coronal and sagittal views to obtain multiple views.
In addition, attention mechanism has achieved excellent results on various computer vision tasks~\cite{ruan2021color,ruan2021dae}.
Along this line, we further employ attention mechanism to identify the most relevant slices in each view, assigning different weights to different slices. 
To accomplish this, we apply the SE module proposed by Hu et al.\cite{hu2018squeeze}, which is typically used to obtain channel attention weights in 2D CNN. 
 In our case, we treat the slices as channels, and the SE module is used to obtain the slice attention weights. 
Inspired by this approach, we propose a Multi-View Attention Learning (MuVAL) method to obtain slice attention weights from multiple views. 

To construct multi-view slice features, we first apply global average pooling to different dimensions of 3D CT images, which can be formulated as follows: 
\begin{equation}
\begin{aligned}
    \bm{f}_t(i) = \frac{1}{h \times w}\sum_{j=1}^{h}\sum_{k=1}^{w}\bm{I}(i,j,k),\\
    \bm{f}_c(j) = \frac{1}{d \times w}\sum_{i=1}^{d}\sum_{k=1}^{w}\bm{I}(i,j,k),\\
    \bm{f}_s(k) = \frac{1}{d \times h}\sum_{i=1}^{d}\sum_{j=1}^{h}\bm{I}(i,j,k),\\
\end{aligned}
\end{equation}
where $\bm{f}_t(i)$ denotes the $i$-th element of $\bm{f}_t$, which is generated by shrinking the $i$-th slice of $\bm{I}$ in the transverse view.

The slice features $\bm{f}_t$, $\bm{f}_c$ and $\bm{f}_s$, can be seen as a condensation of the slice spatial information for each view.
To further obtain the slice attention weights, we employ attention modules consisting of a bottleneck with two fully-connected (FC) layers around the ReLU\cite{nair2010rectified} function with a sigmoid activation:
\begin{equation}
\begin{aligned}
    \bm{\alpha}_t  = \sigma(\bm{W}_{t2}\delta(\bm{W}_{t1}\bm{f}_t)),\\
    \bm{\alpha}_c  = \sigma(\bm{W}_{c2}\delta(\bm{W}_{c1}\bm{f}_c)),\\
    \bm{\alpha}_s  = \sigma(\bm{W}_{s2}\delta(\bm{W}_{s1}\bm{f}_s)),
\end{aligned}
\end{equation}
where $\sigma$ refers to the sigmoid activation, $\delta$ refers to the ReLU function, $\bm{W}_{t1} \in \mathbb{R}^{\frac{d}{r} \times d}$ and $\bm{W}_{t2} \in \mathbb{R}^{d \times \frac{d}{r}}$, $r$ is the dimensionality-reduction ratio.

To enhance the representation of 3D CT images, we apply the average to fuse the multi-view slice attention: 
\begin{equation}
\overline{\bm{I}}(i,j,k)  = \frac{\bm{\alpha}_t(i) + \bm{\alpha}_c(j) + \bm{\alpha}_s(k)}{3}\bm{I}(i,j,k).
\end{equation}

\subsection{Residual Disease Prediction}
For the classification stage, we apply the Med3D~\cite{chen2019med3d} which aggregated several 3D medical image datasets and obtained a series of pre-trained ResNet-3D models. This approach has demonstrated impressive results in various medical challenges and datasets.\cite{li2021alzheimer}.

Considering the computational complexity and available data scale, we employ a pre-trained model of Med3D as the backbone, with a 34-layer ResNet-3D structure. For each input $\bm{\overline{I}}$, the backbone outputs $n$-dimensional features:

\begin{equation}
\bm{f} = Backbone(\overline{\bm{I}}), \bm{f} \in \mathbb{R}^n.
\end{equation}

To ultimately generate the classification probabilities for non-R0 and R0, we apply a 2-output FC layer with a softmax activation:
\begin{equation}
\bm{p} = \sigma(\bm{W}\bm{f}),
\end{equation}
where $\sigma$ refers to the softmax activation, $\bm{W} \in \mathbb{R}^{2 \times n}$, $\bm{p}$ is the one-hot representation for the classification probabilities.

\subsection{Model Learning}
For model learning, 
we employ the cross-entropy as the loss function since it is a classification problem.
The loss function for the output of the last layer is shown as follows:
\begin{equation}
L = -\frac{1}{N}\sum_{i=1}^{N}\bm{y}_ilogP(\bm{p}_i|\bm{I}),
\end{equation}
where $\bm{y}_i$ is the one-hot representation for the true class of the $i^{th}$ instance, and $N$ represents the number of training instances.
\section{Experiment}
\label{s:experiment}
\subsection{Data Description}

In this study, we analyzed 111 patients who were diagnosed with advanced ovarian cancer between March 2018 and June 2020, and whose diagnosis was confirmed through pathology. All patients underwent cytoreduction surgery performed by the same surgical team. As shown in Table~\ref{tab:dataset}, the surgical results for 71 patients were R0, while for the remaining 40 patients, they were non-R0. We randomly selected 88 patients out of the 111 (80\%) for the training set, and 23 out of the 111 (20\%) for the test set. Before cytoreduction surgery, each patient received an enhanced CT scan of the abdominal cavity. Two experienced doctors selected a series of CT slices that contained the entire region of the disease for each patient. The number of CT slices in each sample ranged from 16 to 136, with an average of 63.53 and a median of 61, and the height and width of all CT slices were 512.
The above dataset is desensitized and collected with the consent of all patients, and the study has been approved by the ethics review committee.

\begin{table}
 \centering
 \caption{Number of samples (slices) in the dataset.}
 \label{tab:dataset}
 \tabcolsep=3pt
 \scalebox{1.1}{
 \begin{tabular}{lccc}
  \toprule
  \  & Training set & Test set & In total\\
  \midrule
    non-R0 & 32 (1,712) & 8 (371) & 40 (2,083) \\
    R0 & 56 (3,776) & 15 (1,193) & 71 (4,969) \\
    In total & 88 (5,488) & 23 (1,564) & 111 (7,052) \\
  \bottomrule
 \end{tabular}
 }
\end{table}

\begin{figure*}[t]
\centering
\begin{center}
\begin{minipage}{16cm}
\begin{tabular*}{16cm}{@{\extracolsep{\fill}}c@{}c@{}c@{\extracolsep{\fill}}}
\includegraphics[height=45mm, width=50mm]{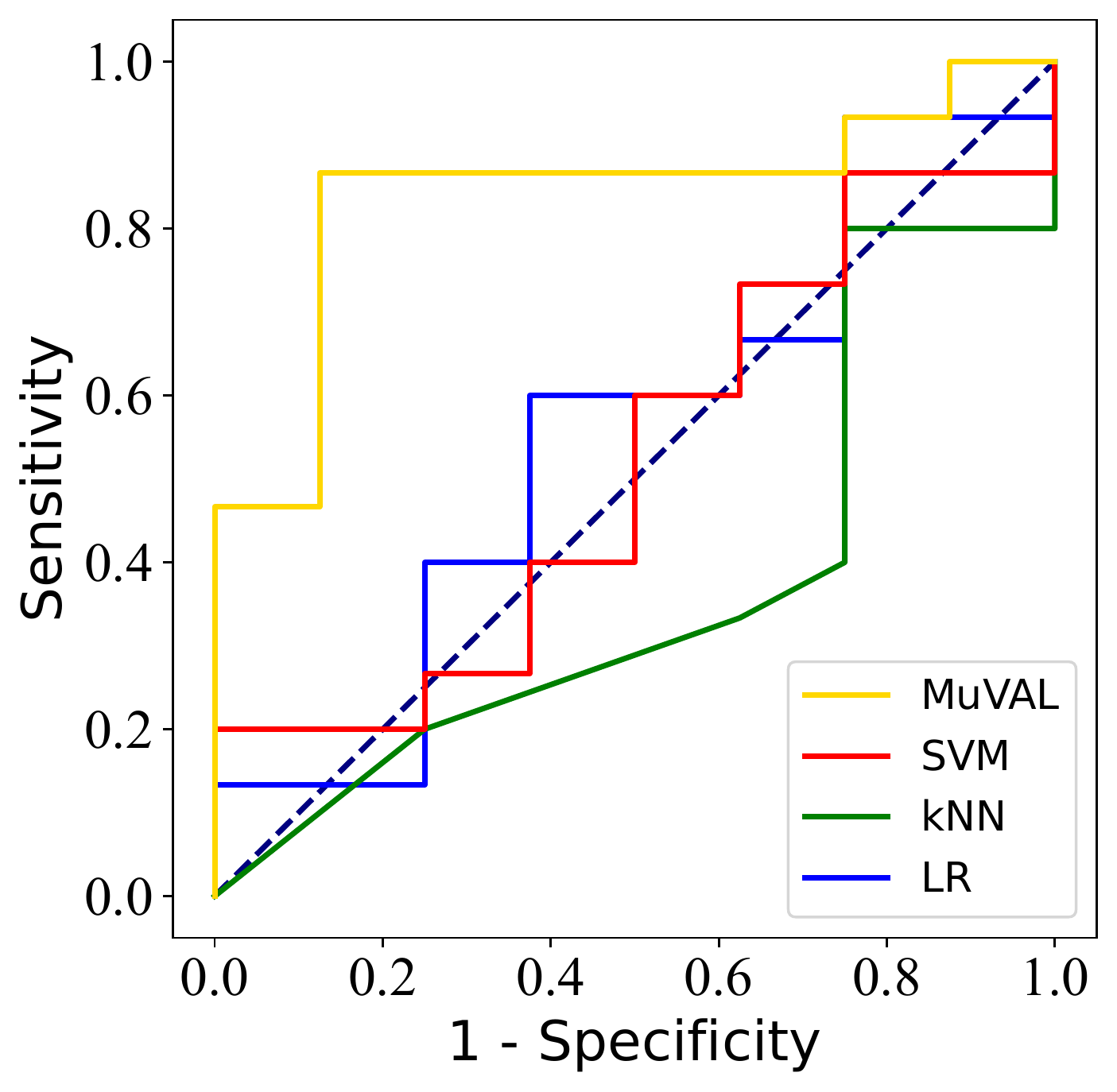} &
\includegraphics[height=45mm, width=50mm]{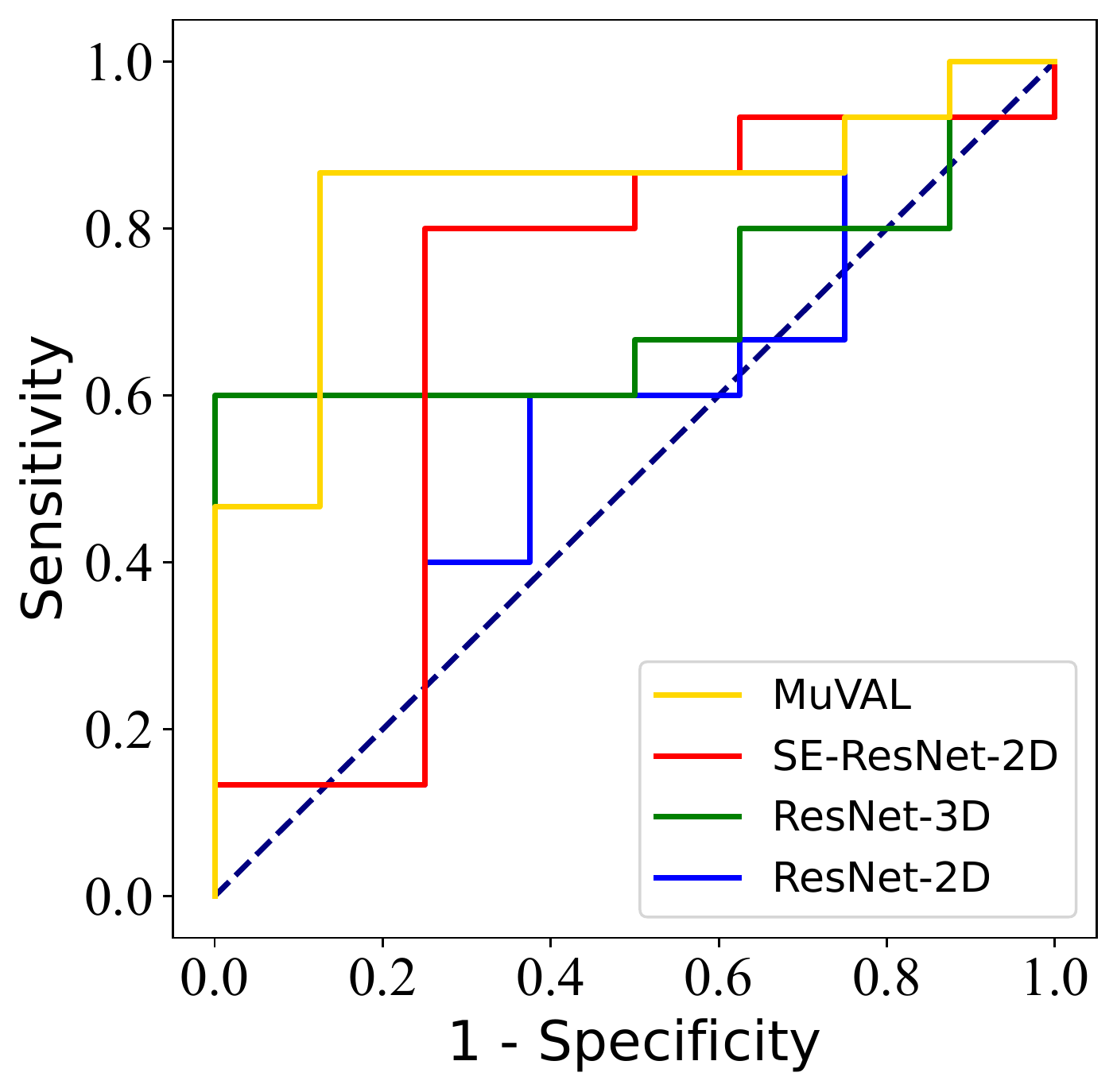} &
\includegraphics[height=45mm, width=50mm]{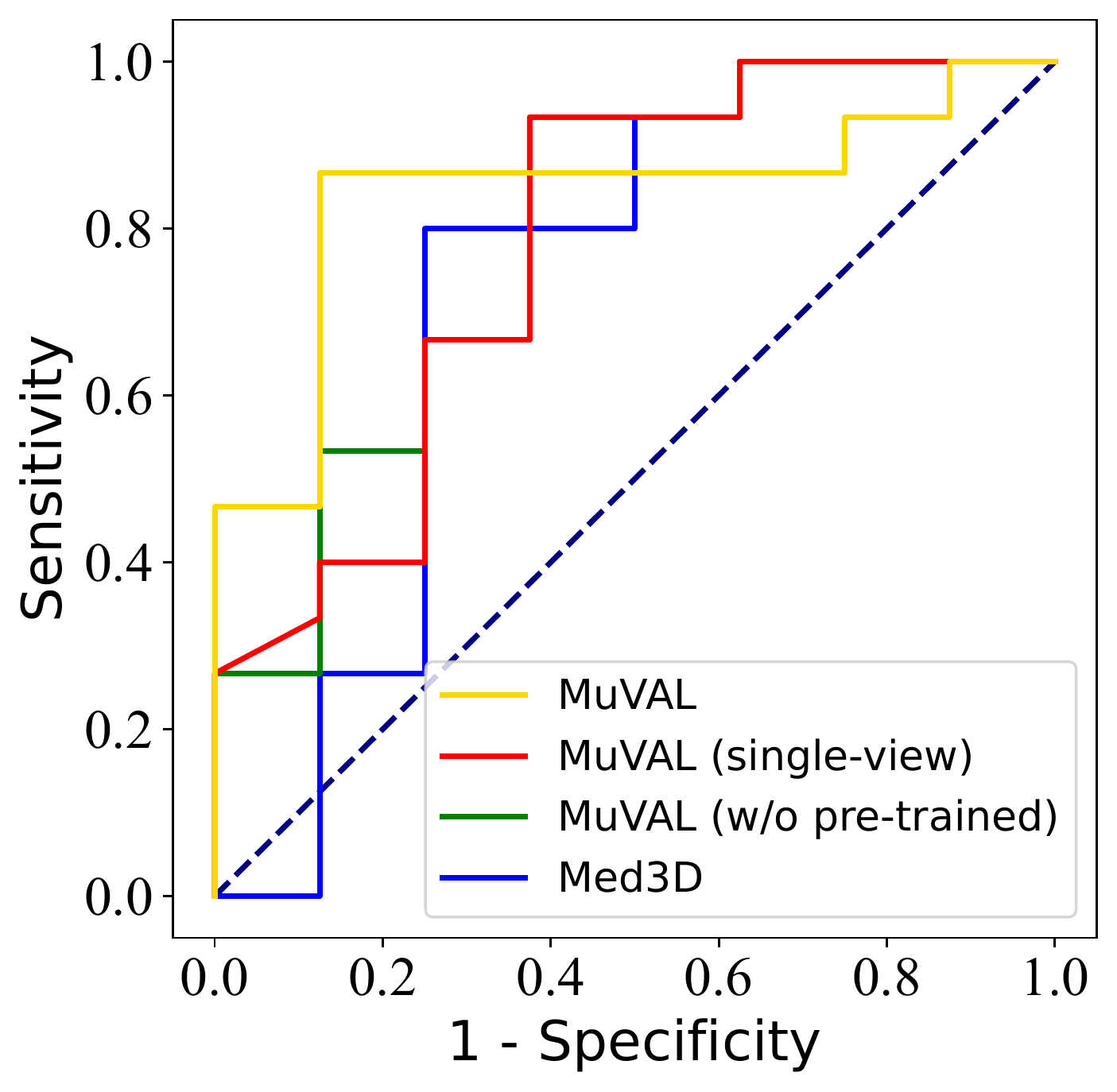} \\
(a) Machine learning methods & (b) deep-learning methods & (c) Ablation performance\\
\end{tabular*}
\end{minipage}
\end{center}
\caption{ROC curves of different methods.}\label{f:roc}
\end{figure*}

\subsection{Experiment Setting}
\begin{itemize}

\item \textbf{Model Setting:} 
In the CT images, the value of each pixel is an integer called CT number, which reflects the attenuation of the X-ray. 
To highlight the ovaries we truncate the CT number to the $[-100, 200]$ interval. 
After truncation, a max-min normalization is applied to map the CT number to $[0, 1]$ interval. 
Considering the mean and median of the number of CT slices, and the computational complexity, we employ bi-linear interpolation to resize all the depth to 64, and the height and width to 256 of all 3D CT images, i.e. $d = 64, h = w = 256$, the dimensionality-reduction ratio r is set to $8$. 
In the classification part, the output dimension of the backbone network $n$ is set to $512$.

\item \textbf{Training Setting:}  
To deal with the small-scale dataset, we apply data augmentation techniques to process the training set, including random translation and rotation, random flipping, and adding Gaussian noise.
In addition to the backbone using pre-trained weights, we set all weights following the Gaussian distribution to initialize the MuVAL model. 
To fine-tune the model, we apply the adamW \cite{loshchilov2017decoupled} to optimize the loss function, and we further initialize the learning rate $lr=0.0001$ and adjust by the exponential decay with $\gamma=0.99$. Moreover, we set the batch size to $10$, the epoch of training to $60$, and utilize an early stop with the patience of $40$. All experiments are implemented on Python 3.8.10, sklearn 1.0.2, Pytorch 1.11.0, and an NVIDIA RTX A5000 GPU with 24GB memory.

\begin{table}
 \centering
 \caption{Results and the number of parameters of models.}
 \label{tab:result}
 \tabcolsep=3pt
 \begin{tabular}{lcccc}
  \toprule
Model & \#params & ACC(\%) $\uparrow$ & F1(\%) $\uparrow$ & AUC $\uparrow$\\
  \midrule
    (1)LR & / &  56.52 & 64.29 & 0.5500\\ %
    (2)kNN \cite{cover1967nearest} & / & 52.17 & 68.57 & 0.3708\\ %
    (3)SVM \cite{cortes1995support} &  / &47.83 & 53.85 & 0.5167\\ %
  \midrule
    (4)ResNet-2D \cite{he2016deep} & 21.47 M & 73.91 & 82.35 & 0.6000 \\ %
    (5)ResNet-3D \cite{he2016deep} & 63.30 M & 73.91 & 75.00 & 0.7000\\ %
    (6)SE-ResNet-3D \cite{hu2018squeeze} & 63.62 M & 78.26 & 82.76 & 0.6917\\
  \midrule
    (7)Med3D \cite{chen2019med3d} & 63.30 M & 78.26 & 82.76 & 0.7250\\
    \makecell[l]{(8)MuVAL\\(w/o pre-trained)} & 63.33 M & 78.26 & 85.71 & 0.7917\\
    \makecell[l]{(9)MuVAL\\(single-view)} & 63.31 M & 82.61 & 87.50 & 0.7792\\ %
    (10)MuVAL & 63.33 M & \textbf{86.96} & \textbf{89.66} & \textbf{0.8417}\\ %
  \bottomrule
 \end{tabular}
\end{table}

\item \textbf{Evaluation Metrics:}
    In general, the Accuracy (ACC) and F1-score (F1) are commonly applied evaluation metrics for classification.
    However, due to the small-scale test set in our task, the ACC may be difficult to compare models with similar performance. 
    In the medical field, the Receiver Operating Characteristic (ROC) curve and Area Under the receiver operating characteristic Curve (AUC) are usually seen as important metrics, which reflect the inter-class distinction and allow comparison of models with similar ACC. 
    Furthermore, ROC and AUC can evaluate the generalization performance of the model in a certain. 
    In conclusion, we combined the above metrics to evaluate the model.
\end{itemize}

\begin{table}
 \centering
 \caption{Case study of MuVAL.}
 \label{tab:case}
 \tabcolsep=3pt
 \begin{tabular}{lccc}
  \toprule
  Sample & Truth & Model prediction & Probability of R0\\
  \midrule
    (1) & R0 & R0 & 88.72\%\\ 
    (2) & R0 & R0 & 84.13\%\\ 
    (3) & R0 & R0 & 86.43\%\\ 
  \midrule
    (4) & non-R0 & non-R0 & 25.68\%\\ 
    (5) & non-R0 & non-R0 & 10.57\%\\ 
    (6) & non-R0 & R0 & 72.71\%\\ 
  \bottomrule
 \end{tabular}
\end{table}

\subsection{Experiment Results} 

\begin{figure*}[t]
\centering
\begin{center}
\begin{minipage}{17.5cm}
\begin{tabular*}{17.5cm}{m{1.5cm}<{\centering} m{2.1cm}<{\centering} m{2.1cm}<{\centering} m{2.1cm}<{\centering} m{2.1cm}<{\centering} m{2.1cm}<{\centering} m{2.1cm}<{\centering}}
 & TP sample (1) & TP sample (2) & TP sample (3) & TN sample (4) & TN sample (5) & FP sample (6)\\
Transverse  &
\includegraphics[width = 2.25cm, height=1.5cm]{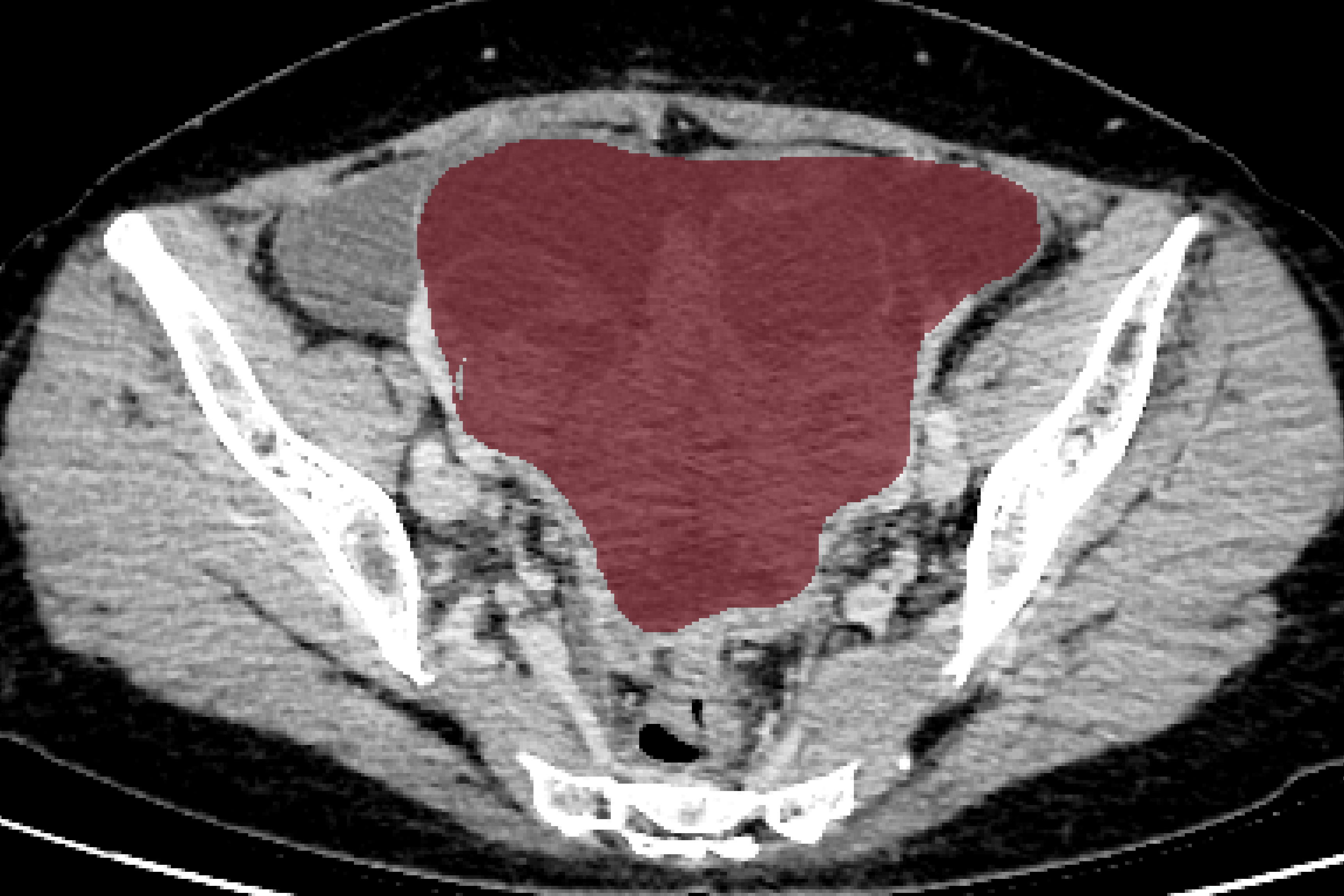} &
\includegraphics[width = 2.25cm, height=1.5cm]{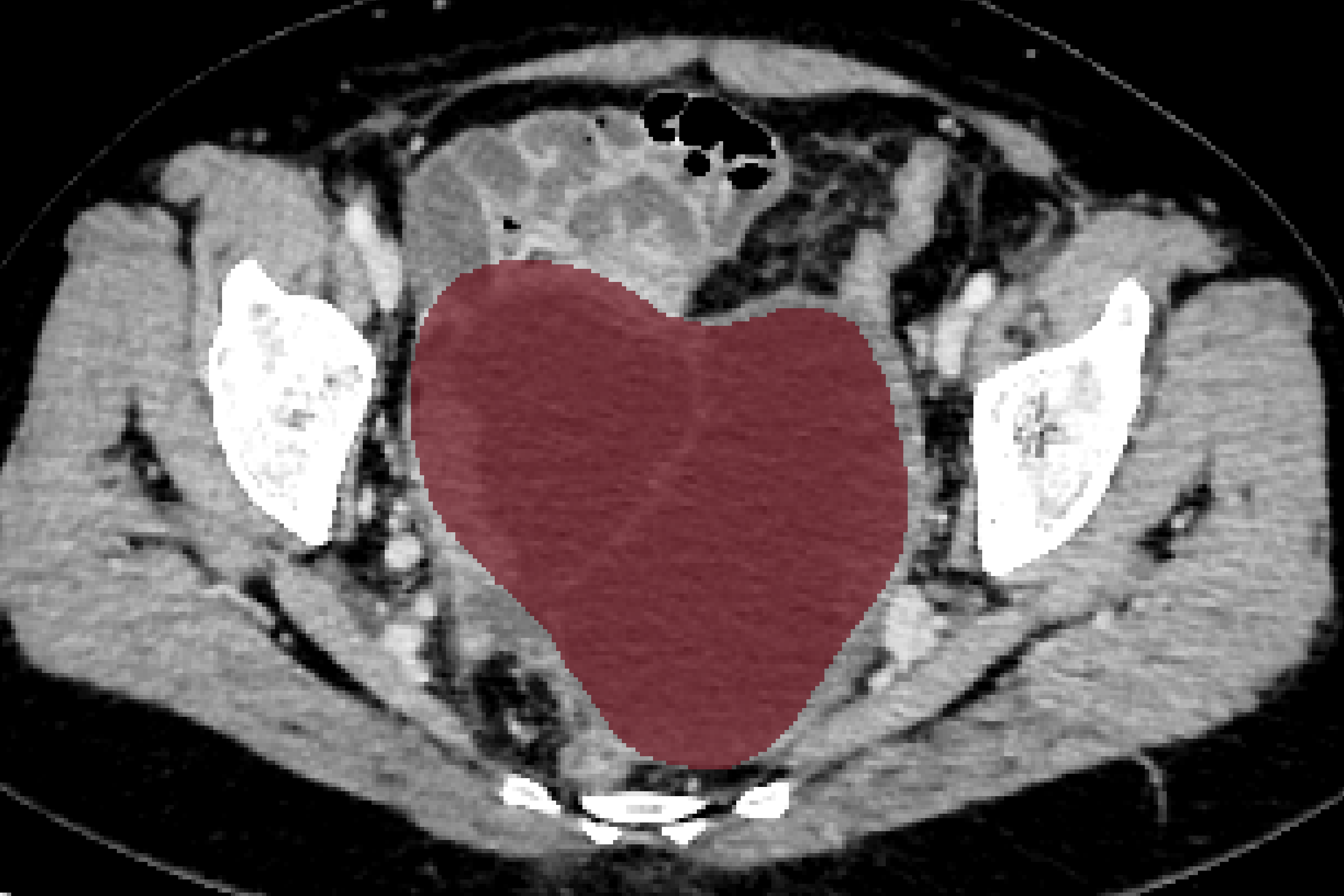} &
\includegraphics[width = 2.25cm, height=1.5cm]{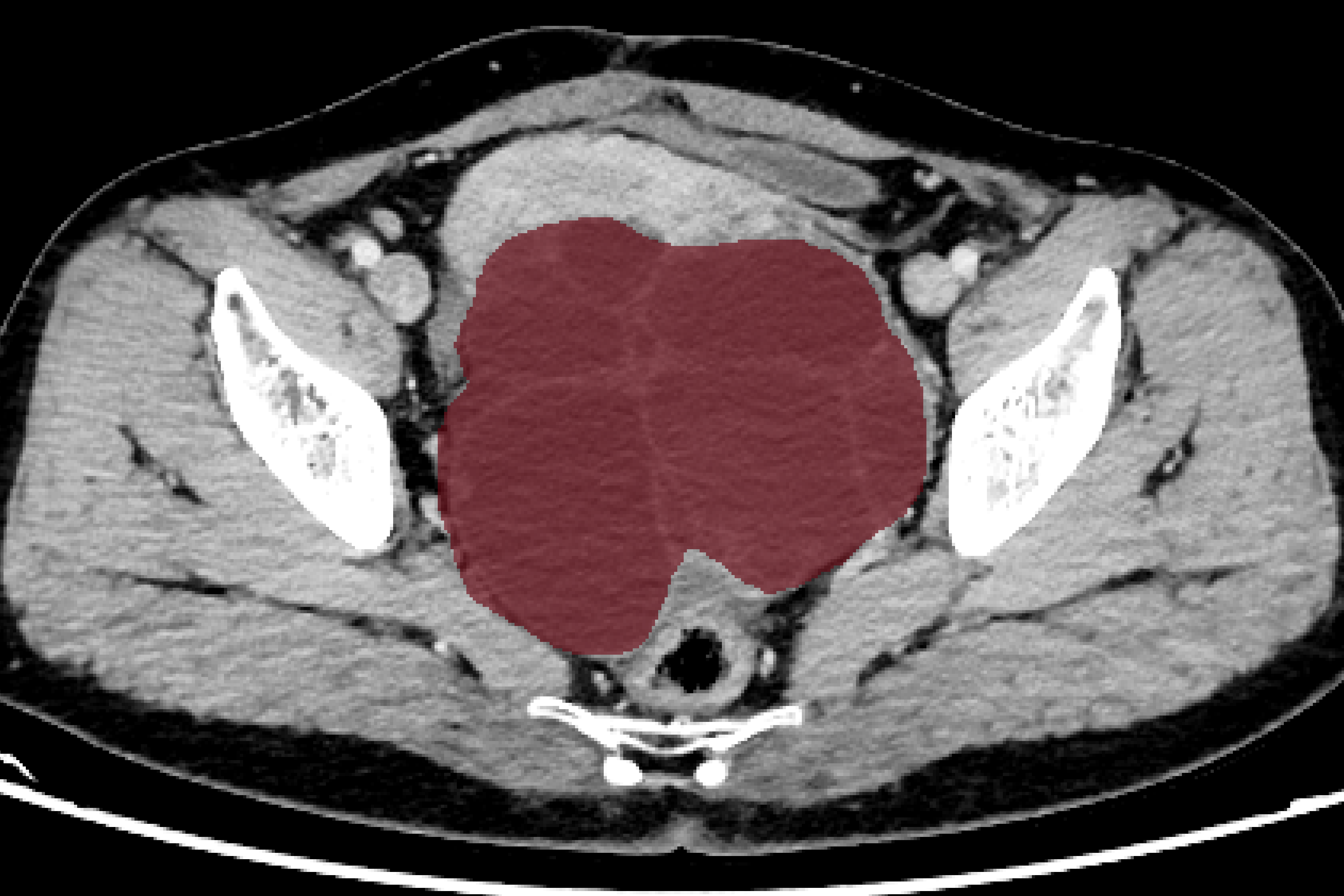} &
\includegraphics[width = 2.25cm, height=1.5cm]{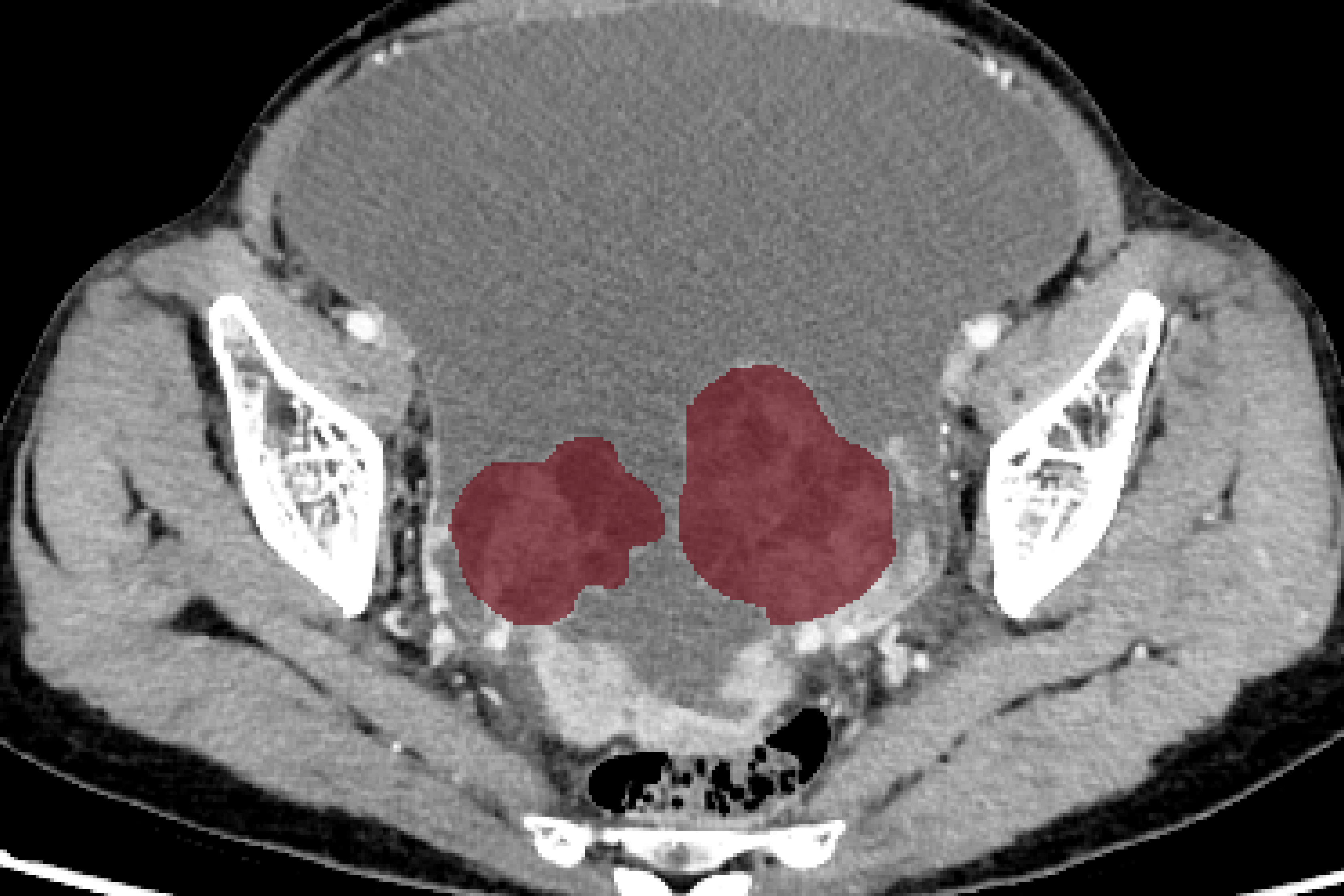} &
\includegraphics[width = 2.25cm, height=1.5cm]{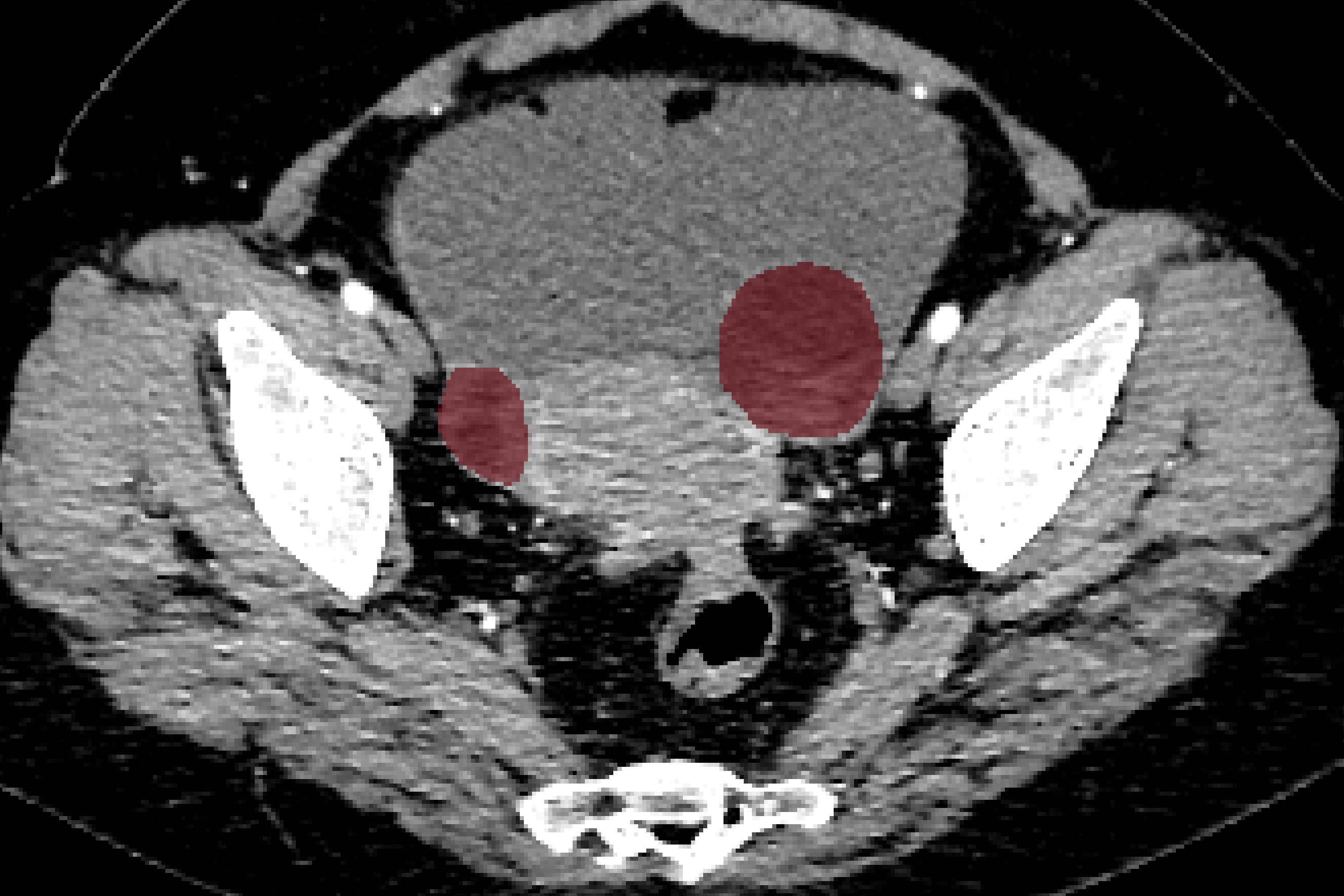} &
\includegraphics[width = 2.25cm, height=1.5cm]{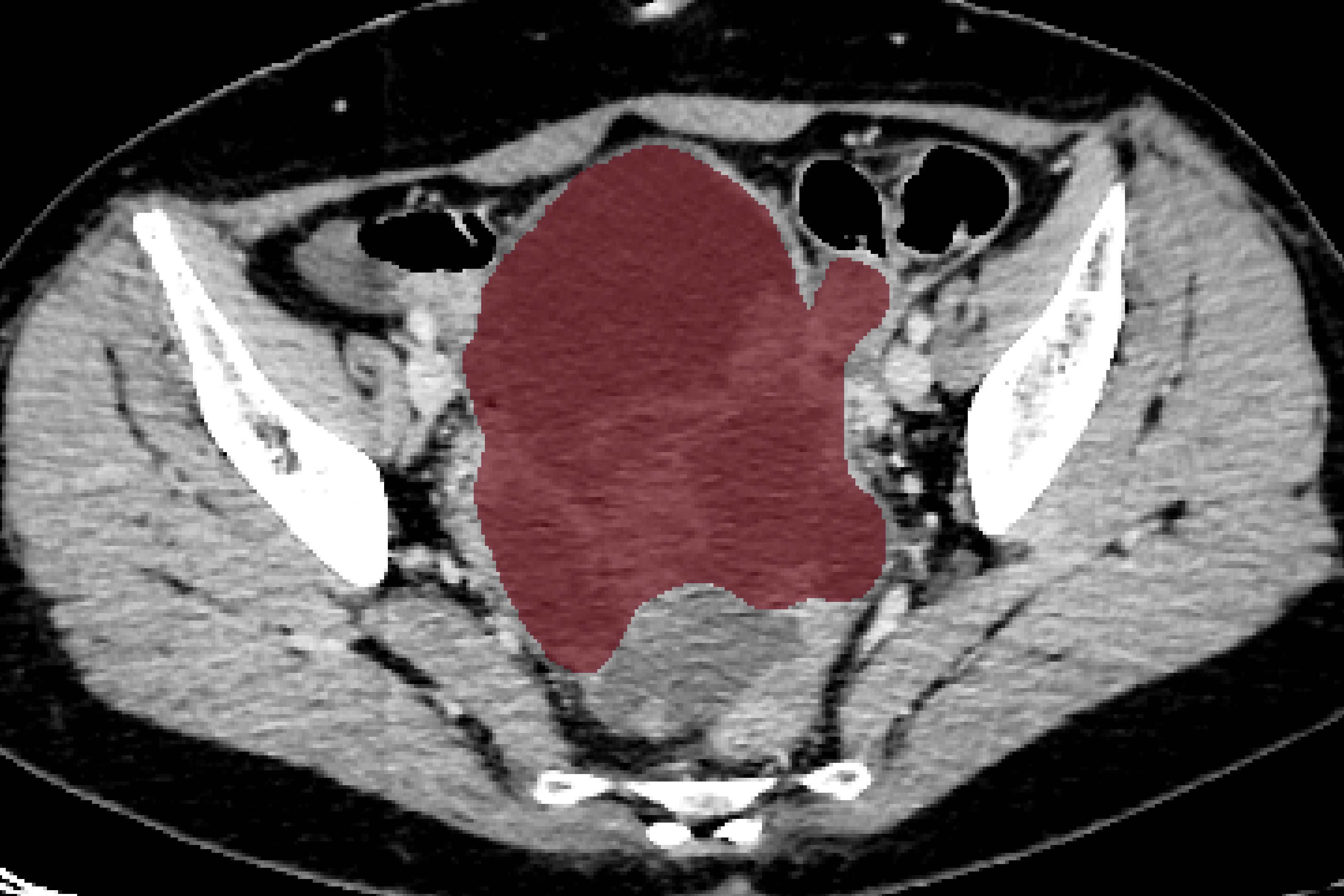} \\
Coronal &
\includegraphics[width = 2.25cm, height=0.5cm]{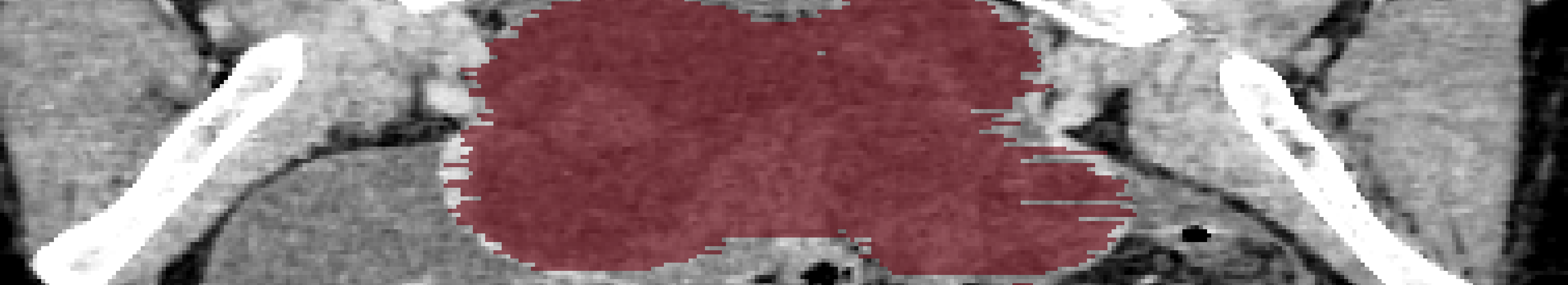} &
\includegraphics[width = 2.25cm, height=0.5cm]{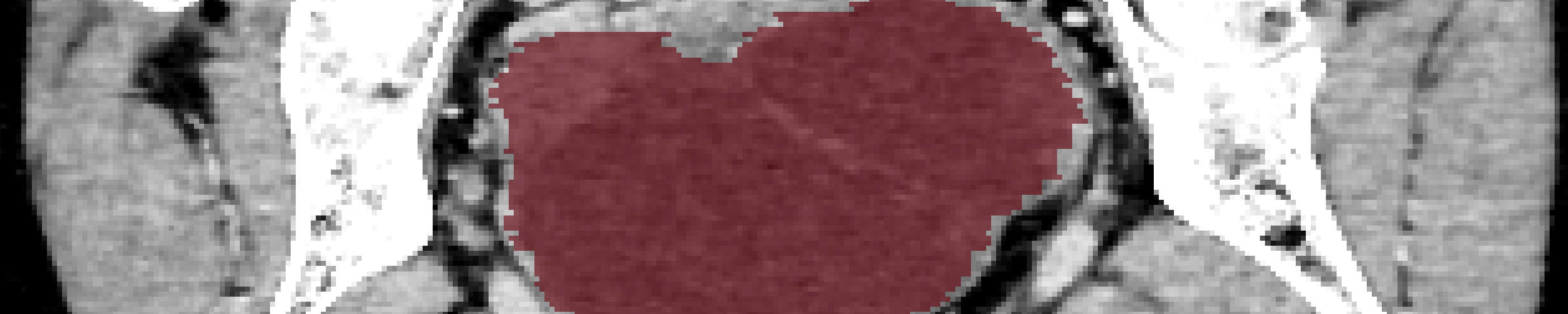} &
\includegraphics[width = 2.25cm, height=0.5cm]{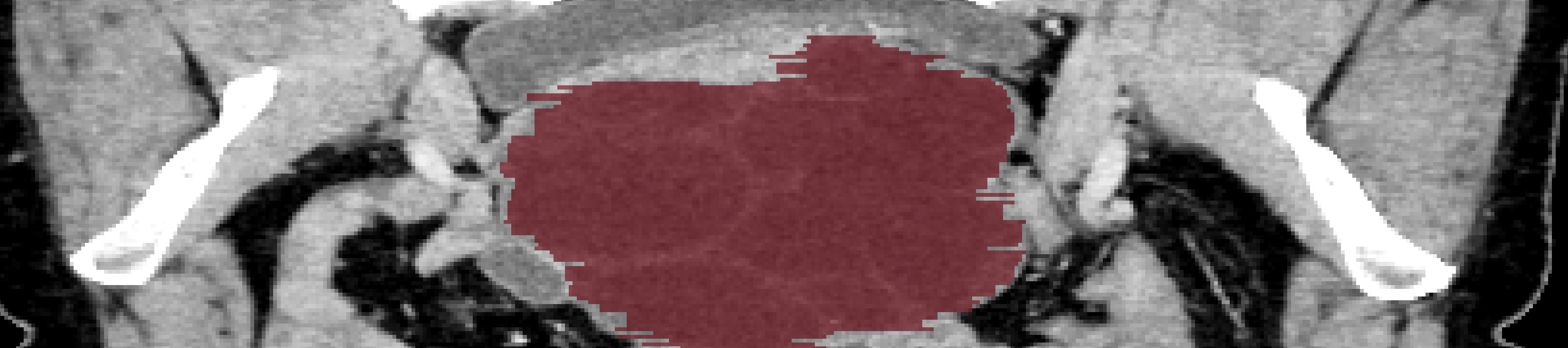} &
\includegraphics[width = 2.25cm, height=0.5cm]{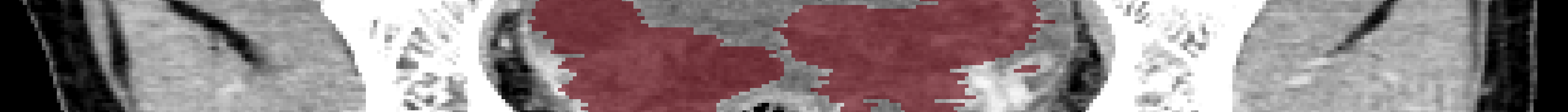} &
\includegraphics[width = 2.25cm, height=0.5cm]{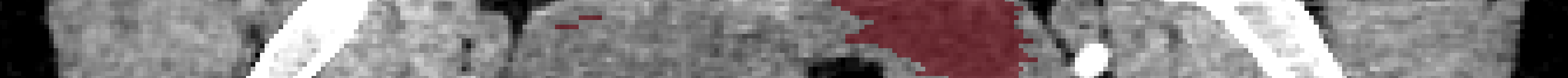} &
\includegraphics[width = 2.25cm, height=0.5cm]{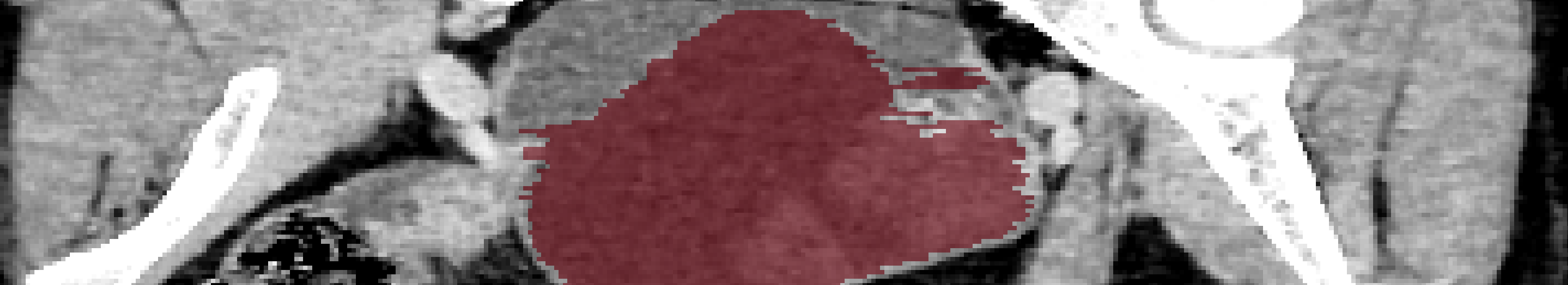} \\
Sagittal &
\includegraphics[width = 2.25cm, height=0.5cm]{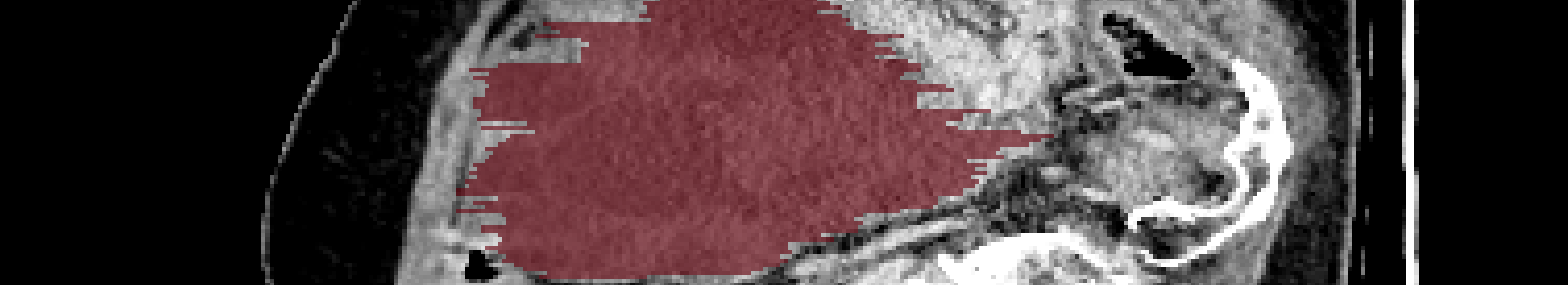} &
\includegraphics[width = 2.25cm, height=0.5cm]{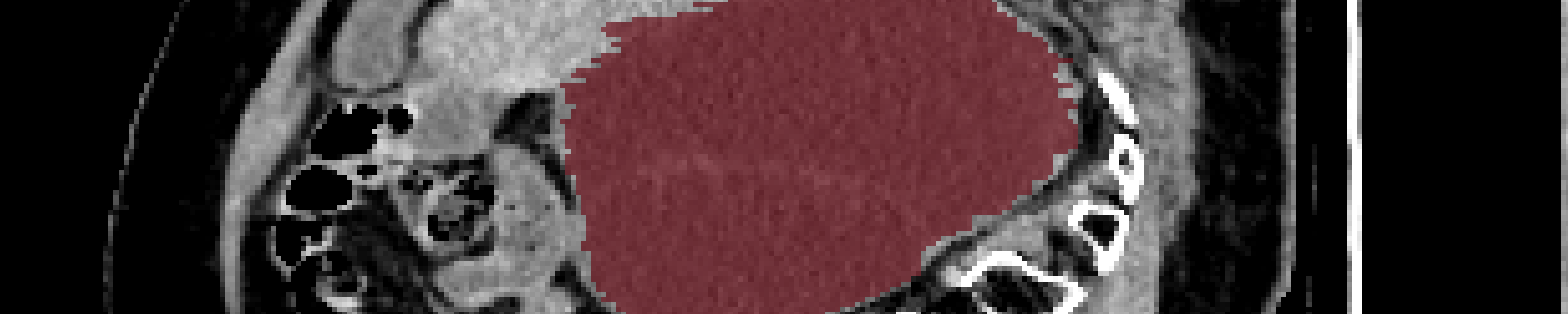} &
\includegraphics[width = 2.25cm, height=0.5cm]{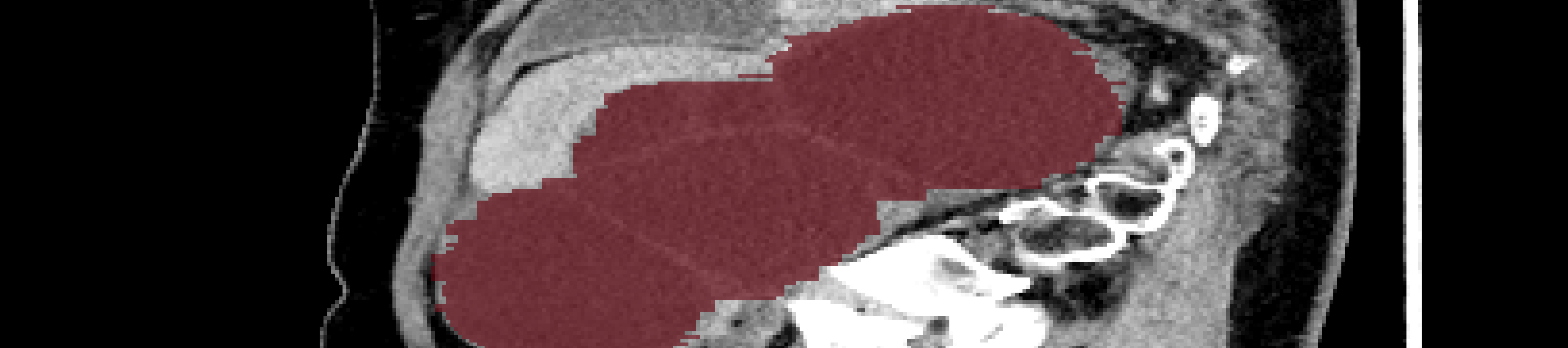} &
\includegraphics[width = 2.25cm, height=0.5cm]{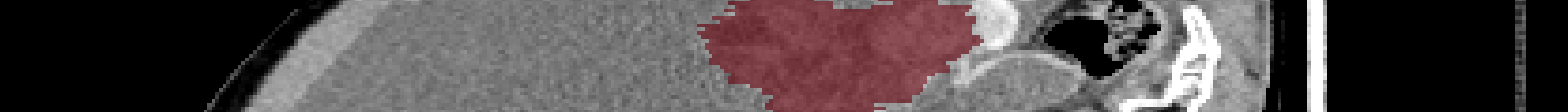} &
\includegraphics[width = 2.25cm, height=0.5cm]{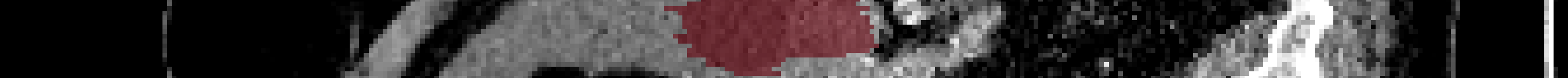} &
\includegraphics[width = 2.25cm, height=0.5cm]{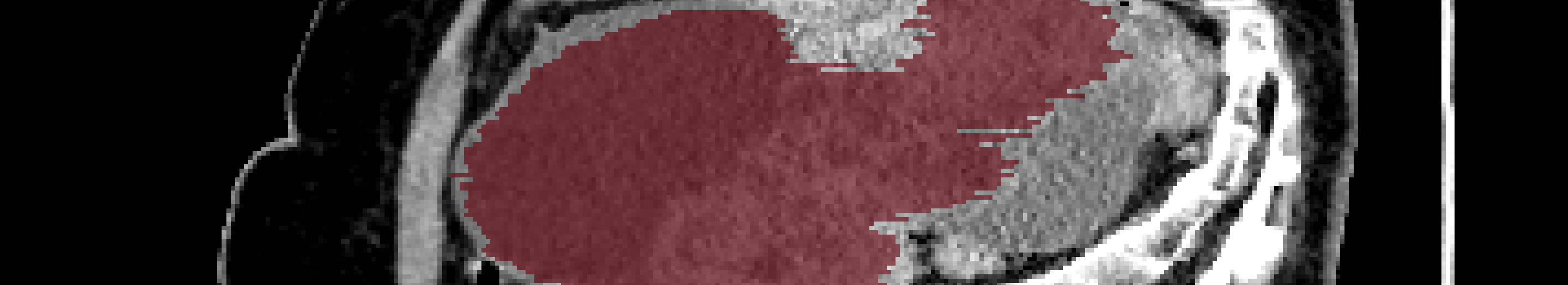} \\
\end{tabular*}
\end{minipage}
\end{center}
\caption{The multi-view slices of samples in Table~\ref{tab:case}, with the disease region outlined. We can observe that positive samples have larger disease regions, while negative samples have smaller disease regions and multiple disease regions.}\label{f:case}

\end{figure*}

We compare our method against the following baselines, Logistic Regression (LR), k-NearestNeighbor (kNN) \cite{cover1967nearest}, Support Vector Machine (SVM) \cite{cortes1995support}, ResNet-2D and ResNet-3D \cite{he2016deep}, SE-ResNet-3D \cite{hu2018squeeze}. All machine learning methods apply grayscale histograms as input, and all ResNets have a 34-layer structure. The overall performance is summarized in Table~\ref{tab:result}, and the ROC curves are shown in Fig~\ref{f:roc}. We can observe that MuVAL achieves the best performance, especially the high AUC indicates that our method has the better inter-class discriminatory ability.

As with traditional machine learning methods, experiments (1)-(3) fail in the task, with all metrics disappointing. Neither of them could well represent image information, especially 3D image information.

Compared to traditional machine learning methods, deep-learning methods, experiments (4)-(6), achieve a high improvement. Specifically, ResNet-2D and ResNet-3D achieve the same ACC. However, ResNet-2D has higher F1 while ResNet-3D has higher AUC, indicating that ResNet-2D has better classification on positive samples, while ResNet-3D has better inter-class distinction. Theoretically, ResNet-3D is more suitable for processing 3D CT images than ResNet-2D. But ResNet-3D has a larger number of parameters, making training more difficult, especially on small-scale datasets, which may result in a similar performance to ResNet-3D and ResNet-2D. With the addition of the attention mechanism, SE-ResNet-3D achieves higher ACC, similar F1 as ResNet-2D, and similar AUC as ResNet-3D. 

\subsection{Ablation Performance}
As shown in Table~\ref{tab:result}, we conduct an ablation study, i.e., experiments (7)-(9).
To compare the performance of the attention part and the classification part, in experiment (7), we remove the entire attention part, remaining only the Med3D backbone.
Similarly, in experiment (8), we remove the Med3D backbone and employ a ResNet-3D without pre-trained instead. 
Even though the results show that the ACC for both is the same, the MuVAL method is more important than the pre-trained backbone because MuVAL without pre-trained achieves higher F1 and AUC.

In experiment (9), we take out the multi-view mechanism, leaving only the transverse view. 
The next-highest ACC and F1 are attained by the single-view model, indicating the performance boost from fusing the attention mechanism with the pre-trained backbone.
However, the large performance gap with the full MuVAL demonstrates the importance of the multi-view mechanism.

\subsection{Case study}
To further analyze the performance of MuVAL, Table~\ref{tab:case} shows the classification results of some test samples. 
According to the prediction and probability, our method performs well on the true positive (TP) samples (1)-(3) and the true negative (TN) samples (4) and (5) but not on the false positive (FP) sample (6).
The multi-view slices of these samples are shown in Fig.~\ref{f:case}, with the disease region outlined. We can observe that the true positive (TP) samples (1)-(3) have larger disease regions, while true negative (TN) samples (4) and (5) samples have smaller disease regions and multiple disease regions. 
The false positive (FP) sample (6) has a large disease region,  which may lead to misclassification. 
The above observations show that our method achieves classification by representing the 3D CT images in terms of disease shape and size. 
However, there may be latent or hard-to-represent information in the 3D CT images such as disease invasion, metastasis, and other information outside the abdominal cavity.
Finally, the results show that a large disease region is more likely to be R0.
The medical explanation \cite{horvath2013relationship} is that the early-stage disease is larger and more complete, while the late-stage disease rapidly detaches and metastasizes widely, causing more surgical difficulties.

\subsection{Discussion}
Despite the effectiveness of our method, several issues remain to be further explored in the future. 
At first, we use a small-scale real-world dataset during the experiments.
Therefore, our method needs to be further tested and evaluated on large-scale datasets. 
Secondly, our case study shows that 3D CT images may be difficult to represent disease invasion and metastasis.
The following research is to apply multi-modal data to improve the performance further. 
Finally, our method may have the potential to predict the residual disease of other cancers, which requires more experiments to verify.
\section{Conclusion}
\label{s:conclusion}

In this paper, we proposed a novel multi-view attention learning method for residual disease prediction of ovarian cancer. Our core idea is to find the more relevant slices in multiple views and to enhance the representation of 3D CT images.
An attention mechanism is applied for each view to obtain the slice attention weights.
As an emphasis, we are the first to introduce a deep-learning method for residual disease prediction of ovarian cancer.
Extensive experiments on the real-world dataset of 111 patients with advanced ovarian cancer demonstrate the superiority of our proposed method.
In particular, the results indicate that our method has improved inter-class discriminatory capacity.

\bibliographystyle{IEEEtran}
\bibliography{ref}

\end{document}